\documentclass[11pt,twoside]{article}

\usepackage{asp2006}
\usepackage{epsf}
\usepackage{psfig}
\usepackage{lscape}

\begin{document}
\title{Anomalous HI Gas in NGC 4395: Signs of Gas Accretion}   
\author{George Heald (1) and Tom Oosterloo (1,2)}   
\affil{(1) ASTRON, Dwingeloo, Netherlands (2) Kapteyn Astronomical Institute, Groningen, Netherlands}    

\begin{abstract} 
In recent years, it has become clear that large quantities of gas reside in the halos of many spiral galaxies. Whether the presence of this gas is ultimately a consequence of star formation activity in the disk, or accretion from outside of the galaxy, is not yet understood. We present new, deep H\,\textsc{i} observations of NGC 4395 as part of a continuing observational program to investigate this issue. We have detected a number of gas clouds with masses and sizes similar to Milky Way HVCs. Some of these are in regions without currently ongoing star formation, possibly indicating ongoing gas accretion.
\end{abstract}

\section{Observations and Data Reduction}

In the study of gaseous halos, most of the recent observational focus has been on nearby normal spiral galaxies \citep[e.g.,][]{o07,b07}. Relatively little attention has so far been paid to low-luminosity spirals, but an H\,\textsc{i} halo has been detected by \citet{mw03} in the edge-on UGC 7321. In order to determine the origin of gaseous halos, it is important to pursue the line of investigation into galaxies with very low rates of star formation, where the effects of star formation and accretion may be more easily distinguished.

To address this issue, we have recently performed deep H\,\textsc{i} observations of the nearby dwarf spiral NGC 4395. The data were obtained using the Westerbork Synthesis Radio Telescope (WSRT) for a total of $8\times12\,\mathrm{hr}$. The target was chosen because it is nearby [$D\,\sim\,3.5\,\mathrm{Mpc}$; \citet{s99}]; viewed at a moderately face-on inclination [$i\,\sim\,46^{\circ}$; \citet{s99}]; isolated; and has a low optical luminosity, with localized star formation activity (see Fig. \ref{fig:anomgas}b).

The data were reduced using standard techniques in the \texttt{MIRIAD} software package. Subsequent analysis was performed using \texttt{GIPSY}.

\section{Data Analysis \& Conclusions}

Inspection of the H\,\textsc{i} data cube reveals the presence of a large population of gas clouds at velocities which place them outside of the normal disk rotation. In total, approximately $5\,\times\,10^7\,M_{\odot}$, equivalent to 5\% of the total H\,\textsc{i} mass, is found to be in this component. Here, we restrict ourselves to presenting a brief overview of the anomalous gas population; we defer a more detailed analysis of the individual clouds to a forthcoming paper.

In order to distinguish the gas at anomalous velocities from the gas participating in normal disk rotation, a velocity field was constructed by tracing the peaks of the individual velocity profiles. Next, the velocity profiles in each line of sight were shifted by the corresponding value in the velocity field, yielding a `derotated' data cube. In this cube, the peak H\,\textsc{i} emission is located (by construction) at $V_{\mathrm{dr}}\,=\,0\,\mathrm{km\,s}^{-1}$. Following a visual inspection of the derotated cube, the gas outside of the somewhat conservative velocity range $|V_{\mathrm{dr}}|\,\leq\,30\,\mathrm{km\,s}^{-1}$ was called anomalous (cf. the typical H\,\textsc{i} velocity dispersion, $7\,\mathrm{km\,s}^{-1}$).

\begin{figure}[!ht]
\plottwo{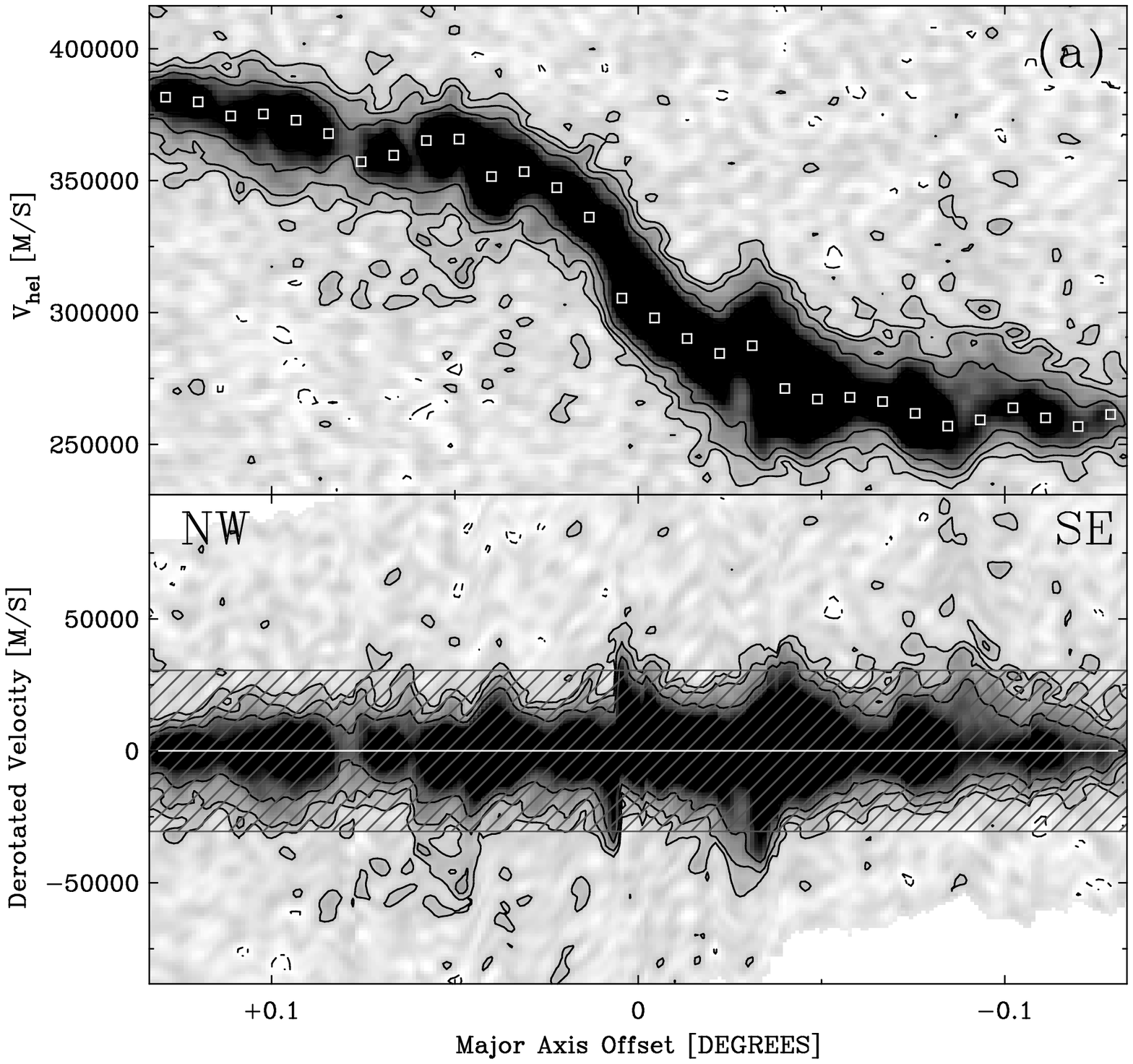}{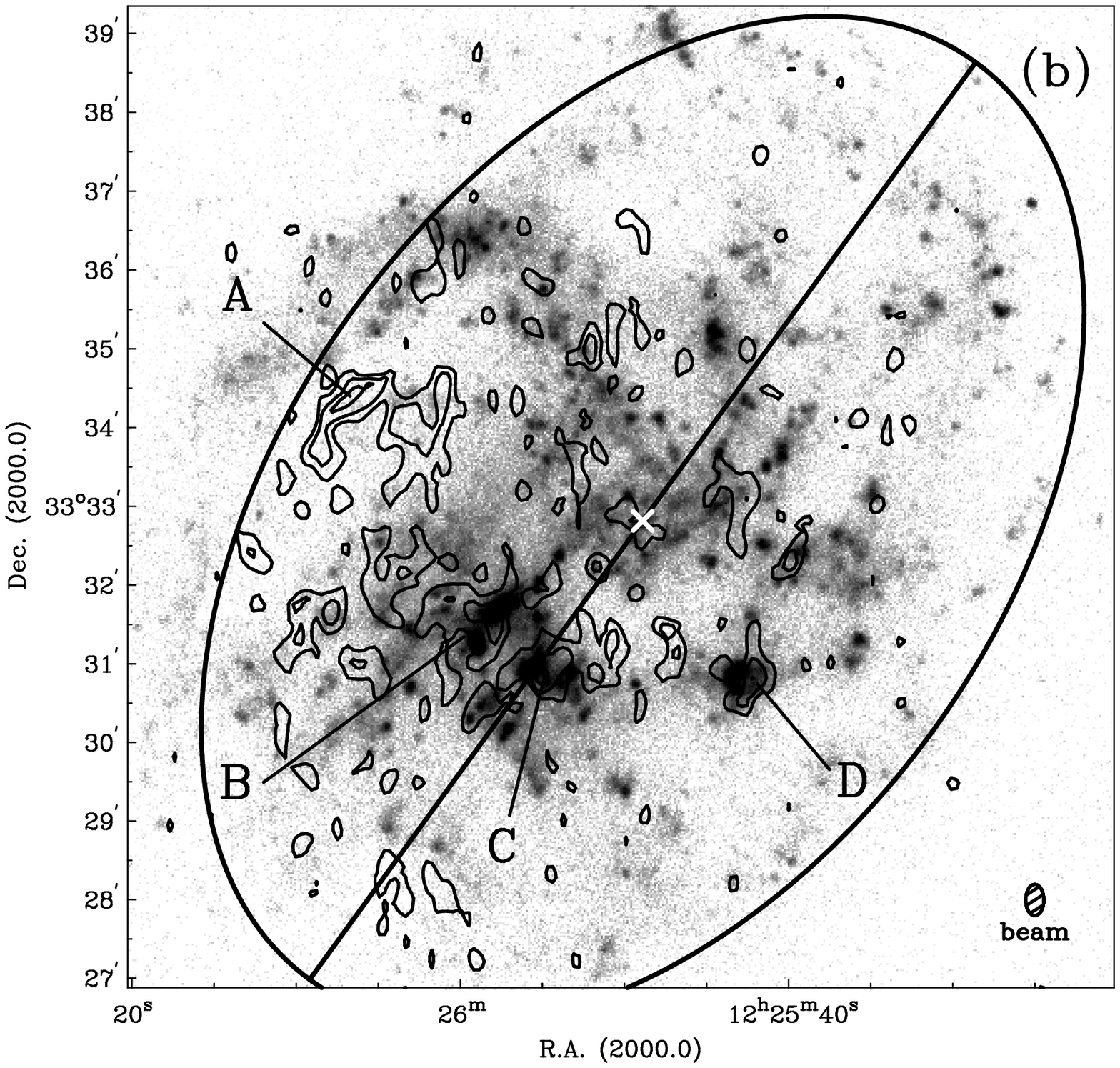}
\caption{(a) Top: Major axis PV diagram; the velocity field values are displayed with white squares. Bottom: Same PV diagram, but derotated. The gas outside of the hatched region is `anomalous'. The contours start at $0.56\,\mathrm{mJy\,beam^{-1}}$ and increase by multiples of 2. (b) Global distribution of the anomalous gas overlaid on the GALEX FUV image. The contours start at $N_{\mathrm{HI}}=1\times10^{20}\,\mathrm{cm^{-2}}$ and increase by multiples of 1.6. The long diagonal line shows the slice along which the PV diagram in (a) was extracted. The white cross marks the rotation center. The ellipse marks $D_{25}\,=\,14.4$ arcmin.}
\label{fig:anomgas}
\end{figure}

The global distribution of the anomalous velocity gas identified using the technique described above is displayed in Fig. \ref{fig:anomgas}b. Four of the larger gas complexes have been labeled (A--D). Complexes B ($M_{\mathrm{HI}}\,\approx\,1.9\,\times\,10^6\,M_{\odot}$), C ($M_{\mathrm{HI}}\,\approx\,2.3\,\times\,10^6\,M_{\odot}$), and D ($M_{\mathrm{HI}}\,\approx\,1.1\,\times\,10^6\,M_{\odot}$) appear to be colocated with regions of active star formation, as suggested by the emission in the underlying GALEX FUV map. Complex A ($M_{\mathrm{HI}}\,\approx\,4.0\,\times\,10^6\,M_{\odot}$) on the other hand, appears to be unassociated with any of the actively star forming regions. It may be a signature of ongoing gas accretion onto the disk of NGC 4395.

\acknowledgements 
The Westerbork Synthesis Radio Telescope is operated by ASTRON (Netherlands Foundation for Research in Astronomy) with support from the Netherlands Foundation for Scientific Research (NWO).


\begin{thebibliography}{}
\bibitem[Boomsma(2007)]{b07} Boomsma, R. 2007, Ph.D. Thesis
\bibitem[Matthews \& Wood(2003)]{mw03} Matthews, L.~D. \& Wood, K. 2003, ApJ, 593, 721
\bibitem[Oosterloo et al.(2007)]{o07} Oosterloo, T., Fraternali, F., \& Sancisi, R. 2007, AJ, 134, 1019
\bibitem[Swaters(1999)]{s99} Swaters, R.~A. 1999, Ph.D. Thesis
\end{thebibliography}
\end{document}